# Decreasing the uncertainty of atomic clocks via real-time noise distinguish

Richang Dong[1,2], Jinda Lin[1], Rong Wei[1*], Wenli Wang[1], Fan Zou[1,2], Yuanbo Du[1,2†],
Tingting Chen[1,2] and Yuzhu Wang[1]

[1]*Key Laboratory for Quantum Optics, Shanghai Institute of Optics and Fine Mechanics, Chinese Academy of Sciences, Shanghai 201800, People's Republic of China*

[2]*University of Chinese Academy of Sciences, Beijing 100049, People's Republic of China*

[†]*Present address: MOE Key Laboratory of Fundamental Physical Quantities Measurement, School of Physics, Huazhong University of Science and Technology, 1037 Luoyu Road, Wuhan 430074, People's Republic of China*

The environmental perturbation on atoms is the key factor restricting the performance of atomic frequency standards, especially in long term scale. In this letter, we demonstrate a real-time noise distinguish operation of atomic clocks. The operation improves the statistical uncertainty by about an order of magnitude of our fountain clock which is deteriorated previously by extra noises. The frequency offset bring by the extra noise is also corrected. The experiment proves the real-time noise distinguish operation can reduce the contribution of ambient noises and improve the uncertainty limit of atomic clocks.

Atomic clocks are being used regularly for timing applications [1-10], precise tests of fundamental symmetries [11-12], search for dark matter [13,14] and quantum information science [15,16]. Caesium fountains have been contributing to TAI as the most accurate and precise primary frequency standards for more than a decade [17,18], and rubidium fountains has been accepted as a secondary standard [6,8]. To date, atomic fountain clocks (AFCs) have achieved total uncertainty of a low $10^{-16}$ level and short term stability of low $10^{-13}\tau^{-1/2}$ to $10^{-14}\tau^{-1/2}$ [6-9]. Optical lattice clocks and ion clocks have progressed with a rapid pace, and now achieved a lower systematic uncertainty of $10^{-18}$ [1-4,19].

Several common physical effects induced by the ambient perturbations, like the blackbody radiation (BBR) effect, the ac Stark effect, and the density effect, *etc.*, are the major terms that affect the performance of atomic clocks [2,10,20]. Top clock labs have developed amazing techniques to control the environmental fluctuation and to minimize the systematic uncertainties. In order to address the BBR induced frequency shift, NIST-F2 operated at a cryogenic temperature of 80 K [7], Katori's group interrogated Sr atoms of optical clock in a cryogenic environment of 95 K [4], and Ye's group used a moveable temperature sensor to predict the dynamic BBR shift [1,2]. To reduce the collision shift, NPL and NIST operated their atomic fountains at low densities [7,21]. Godun et al. measured the laser power in real time to decrease the fractional uncertainty of the ac Stark shift in Yb$^+$ clock to a level of $8 \times 10^{-18}$ [22]. Real-time monitor of the related environmental parameters with high resolution is used to distinguish the correlated noises. The frequency shifts of clock transition transferred from the corresponding noises can be effectively distinguished and subtracted.

In this letter, we propose an effective method based on real-time noise distinguish (RTND) to improve the atomic clock's performance limited by ambient perturbations. A proof-of-principle experimental demonstration is fulfilled in an AFC [23,24]. The magnetic field that related to the 2$^{nd}$ Zeeman effect is deteriorated by an analog noise. At the same time a high precision sensor is used to monitor the fluctuations of magnetic field. Combining the detected environmental noise and the error signal, the deterioration from the analog noises is corrected by RTND, which leads to the improvement of clock stability at longer time scale. The demonstrated innovation can be applied to a variety of different physical effects on various atomic clocks to improve the accuracy and stability.

In an atomic clock, the high stability is realized by referencing the stable oscillator to a high-quality-factor atomic transition profile. In a microwave clock operating with Ramsey sequence, the atomic transition probability is given by $[1+\cos(2\pi\delta\nu T)]/2$. Here $T$ is the interrogation time, $\delta\nu$ is given by

$$\delta\nu = \nu_{LO} - \nu_{AT} \qquad (1)$$

where $\nu_{LO}$ is the frequency of the LO, and $\nu_{AT}$ is the atomic resonance frequency.

With a closed servo loop, the LO frequency $\nu_{LO}$ is locked to the atomic reference frequency $\nu_{AT}$. The superb reproducibility of $\nu_{AT}$ ensures the accuracy of clock output. The two-sample Allan deviation, characterizing the stability of a clock output, averages as $\tau^{-1/2}$ or $\tau^{-1}$ [25]. However, $\nu_{AT}$ is actually not constancy due to the coupling of the atomic internal states from ambient electromagnetic fields, albeit under careful control, then

$$v_{AT} = v_0 + \sum_{i=1}^{N} v_i \quad (2)$$

Here $v_0$ is the atomic resonance frequency without environmental perturbations, $v_i=f_i(x_i)$ is the frequency shift arising from the $i$-th environment related physical parameter $x_i$, $N$ is the total number of related physical parameters. From Eq. (2) with Eq. (1), we obtain

$$\delta v = v_{LO} - (v_0 + \sum_{i=1}^{N} v_i) \quad (3)$$

As illustrated in Fig. 1 (red line servo loop), the measured error signal $\delta v$ containing the environmental noise in the last clock cycle feeds back to the LO, and influences the frequency of LO in the current clock cycle. The environmental fluctuation then degrades the clock stability as well as accuracy. Therefore the performance of the clock, especially at longer time scale, is limited by the environmental noise.

As shown in Fig. 1 (blue line), our RTND approach uses a detection system to measure $x_i$ in real time. Considering the experimental case involving M($\leq$N) measurable physical parameters, with Eq. (3) the corrected frequency error is given by

$$\delta v_{RTND} = (v_{LO} - v_0) - (\sum_{i=1}^{N} v_i - \sum_{j=1}^{M} v_j^D) . \quad (4)$$

Here $v^D_i=f^D_i(x^D_i)$ is the detected frequency shift when operaring in the RTND condition. With high-precision noise detection, $v^D_i$ approximately equals to $v_i$. The RTND approach suppresses the environmental impacts on clock performance and locks the clock output to noise immune atomic resonance frequency.

It is more obvious to analyze the frequency stability of the two schemes. In the standard feedback scheme, the total uncertainty $\sigma_0(\tau)$ is deteriorated by environmental noise, and turns out to be

$$\sigma(\tau) = \sqrt{\sigma_0^2(\tau) + \sum_{i=1}^{N} \sigma_i^2(\tau)}, \quad (5)$$

where $\sigma_i(\tau)$ is the uncertainty of $v_i$ and they are independent. The clock stability is ultimately limited by the environmental noise, especially at longer averaging time. With RTND scheme detecting the ambient fluctuation in real time, the induced frequency errors are eliminated from $\sigma(\tau)$ as

$$\sigma_{RTND}(\tau) = \sqrt{\sigma^2(\tau) - \sum_{i=1}^{M} \sigma_i^2(\tau)}. \quad (6)$$

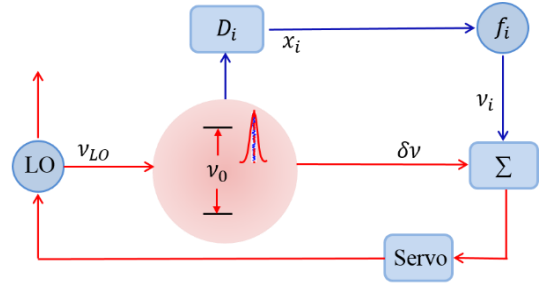

FIG. 1 (color online). Principle schematic of the RTND demonstration. The additional sensors group $D_i$ is used to detect the environment-related physical parameters $x_i$ in real time (blue line).The influence of noise on frequency $v_i=f_i(x_i)$ are subtracted from the error signal $\delta v$.

As a result, the RTND protocol breaks the uncertainty limit on normal locking mode.

We experimentally demonstrate the concept of RTND in our $^{87}$Rb AFC by a noise deterioration method [26]. The extra noise is added to exaggerate the fluctuation of the corresponding parameter and depicts the improvement of RTND. Specifically, the 2$^{nd}$ Zeeman shift is used as the demonstrating effect. The noise, Allan deviation of which is between $10^{-15}$ and $10^{-14}$, mimics the fluctuation of atom number of our AFC. The amplitude of noise is set according to H-maser (VCH 1003A) with noise floor of about $2 \times 10^{-15}$ [23], since the demonstration of RTND is realized by phase comparing AFC with H-maser. In normal AFC mode, the noise deteriorates the performance of the frequency output of AFC, while in RTND mode, the noise is detected and rejected.

As shown in Fig. 2, a solenoid surrounding the flight tube is used for supplying the bias magnetic field with strength of 130.2 nT [27]. The subtle fluctuation of the current value is negligible for a clock with quadratic Zeeman shift of $142.6 \times 10^{-15}$ and long term stability of $1.6 \times 10^{-15}$ [23]. Then the noise values are seriatim added to the C-field solenoid for every clock cycle. A precise resistor with calibrated linearity is used to track the fluctuation of the C-field current. The induced frequency shift is $v_B=f_B(B)=k_Z\langle B\rangle^2=k_C\langle I\rangle^2$, here $B$ is the measured C-field value, $k_Z$ and $k_C$ are the coefficients [27], $\langle I \rangle$ displays the average C-field current on the cycle. Combining the normally measured error signal $\delta v$ and magnetic field induced frequency shift $v_B$, the final error is $\delta v_{RTND}=\delta v+v_B$ with uncertainty

$$\sigma_{RTND}(\tau) = \sqrt{\sigma^2(\tau) - \sigma_B^2(\tau)} . \quad (7)$$

Where $\sigma_B(\tau)$ is the uncertainty of $v_B$.

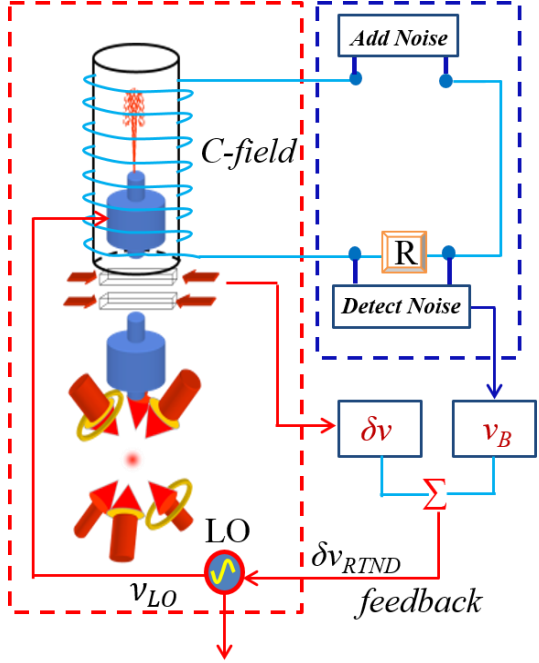

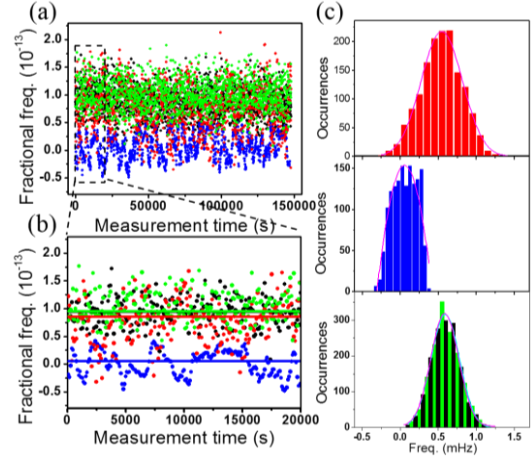

FIG. 2 (color online). Experimental diagram of AFC operating in the RTND protocol. The C-field of AFC (red line box) is deteriorated by adding noise to the current of the solenoid. A sampling resistor is used to detect the fluctuation of current which directly reflects the fluctuation of the magnetic field (blue line box). The influences of the fluctuation are eliminated from the error signal before feedback to LO.

FIG. 3 (color online). Frequency differences between the $^{87}$Rb AFC and the H-maser. (a) Frequency differences for 150,000 s at natural conditions (black dots), C-field noise on (red), and RTND on (green). Blue dots denote the frequency shift corresponding to real-time detected C-field noise. (b) Dataset of frequency difference over a 20,000 s interval. The horizontal lines indicate the mean fractional frequency values of $7.8 \times 10^{-14}$, $8 \times 10^{-15}$, $8.7 \times 10^{-14}$ and $8.6 \times 10^{-14}$ for red, blue, green, and black lines, respectively. (c) Histograms of all data corresponding to red, blue, green (black) dots. The linewidth of green (black) histogram is narrower than of red higstogram.

The fountain clock is running under three different states: 1) no extra noise for magnetic field (natural conditions); 2) adding noise to magnetic field; 3) RTND correction when noise in magnetic field. To facilitate the description, we substitute the frequency of AFC for frequency differences between AFC and H- maser.

The frequency differences of the three states are shown in Fig. 3(a). Each data represents the average result of 100 s, and the total measurement time is 150, 000 s. A group of noises (blue dots) are applied so that the clock's transition frequency is perturbed (red dots). This will decrease the stability in the characteristic averaging time, especially at long time scale. The amplitude of fluctuation of frequency difference after RTND (green dots) is as small as natural conditions (black dots). Therefore, the long term drifts and fluctuation caused by the magnetic field noise should be eliminated. In addition, the frequency offset arising from the noise, shown with the horizontal line in Fig. 3(b), is also corrected by the protocol. Here, the black line denotes the inherent frequency difference between the AFC and H-maser, while the interval between the red line and the black line comes from the noise offset.

Table 1. Frequency offset corrected by RTND ($\times 10^{-13}$).

| Noise offset | Total offset | corrected |
|---|---|---|
| 0 | 0.98 | 0.97 |
| -1 | 1.91 | 1.01 |
| 0.6 | 0.41 | 0.96 |

More data shows that noise with a negative offset of $-1 \times 10^{-13}$ and a positive offset of $0.6 \times 10^{-13}$ are all corrected by RTND, leaving the fractional frequency difference coincident with that of zero offset, as shown in Table 1. In this way, we can compensate the frequency shifts listed in the type-B uncertainty budget using the RTND.

In frequency domain, the histogram statistics of the frequency differences under natural conditions is shown in Fig. 3(c) (black histogram). Carrying noises (blue histogram), the frequency differences (red histogram) shows a wider linewidth than that of natural conditions. After RTND, the distribution of frequency difference (green histogram) is consistent with that under natural conditions.

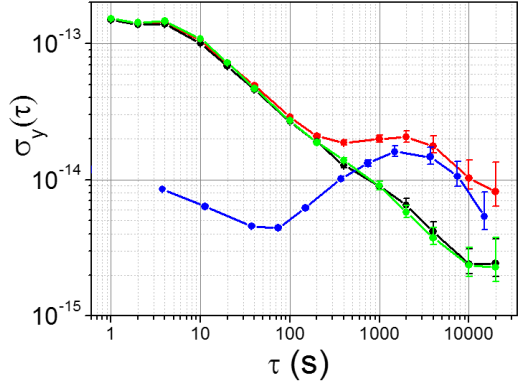

FIG. 4 (color online). Normal Allan deviation of the frequency difference when the AFC run at natural conditions (black), deteriorated by noises (red line), and corrected by RTND (green line). Blue line: Allan deviation of real-time detected noise.

To characterize the performance of the RTND in AFC, we evaluate the stability result in terms of Allan deviation. Under natural condition, the Allan deviation of AFC is $2.7 \times 10^{-13}\tau^{-1/2}$ and reaches $2.5 \times 10^{-15}$ for $\tau = 10,000$ s, as shown in Fig. 4 (black line). The Allan deviation of the extra noise is depicted by blue line, whose value is between $5 \times 10^{-15}$ and $2 \times 10^{-14}$. When the average time is less than 200 s, the instability of the noises is far smaller than the instability of AFC, and it becomes worse as the average time exceed 500 s. After the noises are applied, the performance of the AFC becomes worse when the averaging time is larger than 200 s (red line). The final result of Allan deviation follows the Eq. (5) in normal clock locking scheme. After RTND, the instability of AFC is reduced (green line) and the result is as good as that under natural condition (black line), which means the uncertainty arising from noises is eliminated. We infer that the RTND method can also effectively suppress some other environmental perturbations on atomic clock.

In conclusion, we demonstrated a real-time noise distinguish method to eliminate the corresponding type-B uncertainties and improve the frequency stability for atomic clock. The influence of the noises is effectively suppressed and the long term stability of atomic clock is improved. By RTND protocol, the improvement of stability and uncertainly of atomic clock lies on accurate measurement of the related parameters and the transfer function. The development of RTND protocol opens up new directions and possibilities to the improvement of accuracy of other clocks, like the Cs fountain clocks and the Sr optical lattice clocks, by inhibit the impact of collisional shift, the BBR shift, the optical lattice effect, *et al.*

Richang Dong and Jinda Lin contribute equally to this work. We thank Professor Bo Yan and Professor Jinming Liu for useful discussions and this work was supported by the National Natural Science Foundation of China under Grant Nos. 61275204 and 91336105.